\documentclass[12pt]{iopart}
\usepackage{graphicx}
\begin{document}

\title[Wilson-Polchinski exact RG equation]{Wilson-Polchinski exact renormalization group equation for $O(N)$ systems:
 Leading and next-to-leading orders in the derivative expansion.\dag\ }

\author{C.\ Bervillier}

\address{Service de physique th\'{e}orique,
CEA/DSM/SPhT-CNRS/SPM/URA 2306
CEA/Saclay,
F-91191 Gif-sur-Yvette C\'{e}dex, France \\ and \\
Laboratoire de Math\'ematiques et Physique Th\'eorique, CNRS/UMR 6083, Universit\'e de Tours,
Parc de Grandmont, 37200 Tours, France}

\begin{abstract}
With a view to study the convergence properties of the derivative expansion
of the exact renormalization group (RG) equation, I explicitly study the
leading and next-to-leading orders of this expansion applied to the
Wilson-Polchinski equation in the case of the $N$-vector model with the
symmetry $\mathrm{O}\left( N\right) $. As a test, the critical exponents $%
\eta $ and $\nu $ as well as the subcritical exponent $\omega $ (and higher
ones) are estimated in three dimensions for values of $N$ ranging from $1$ to $20$. I compare
the results with the corresponding estimates obtained in preceding studies
or treatments of other $\mathrm{O}\left( N\right) $ exact RG equations at
second order. The possibility of varying $N$ allows to size up the
derivative expansion method. The values obtained from the resummation of
high orders of perturbative field theory are used as standards to illustrate
the eventual convergence in each case. A peculiar attention is drawn on the
preservation (or not) of the reparametrisation invariance.
\end{abstract}


\submitto{\JPCM}

\ead{bervil@spht.saclay.cea.fr}

\dag\ Dedicated to Lothar Sch\"afer on the occasion of his 60th birthday


\section{Introduction}
The renormalization group (RG) theory is suitable to the study of many
modern physical problems. Generically, every situation where the scale of
typical physical interest belongs to a (wide) range of correlated or coupled
scales may be (must be?) treated by RG techniques. Critical phenomena, which
are characterized by one (or several) diverging correlation length(s),
provide \textquotedblleft the\textquotedblright\ didactic example \cite{440}.
Quantum field theory, with its strongly correlated quantum fluctuations, is
not less famous since it has given rise to the early stages of the RG theory 
 \cite{206}.

Thanks to a fortunate success (essentially due to an impressive diagrammatic
calculation \cite{323}) in estimating the critical
behavior of some systems \cite{5459,4948}, the perturbative framework has
pushed into the background the undoubtedly nonperturbative character \cite%
{2839,32,5865} of the RG theory. As a consequence, there has been relatively
little interest in the development of nonperturbative RG techniques \cite%
{4595}. In particular the formulation of the RG theory via an infinitesimal
change of the scale of reference (running scale), designated by the generic
expression \textquotedblleft exact renormalization group
equation\textquotedblright \cite{440}\ although known since 1971 \cite{4495},
has actually been actively considered only since the beginning of the
nineties \cite{4595}. Because the variety of systems to which the exact RG
formulation could be applied is large \cite{4595,4700} (see also \cite{5701}%
) and also because the
perturbative framework is generally not well adapted to such studies \cite%
{5865,5744,5745}, it is worthwhile making every endeavour to master, if
possible, the exact RG framework.

The exact RG equation is an integro-differential equation the study of which
calls for approximations and/or truncations. Among the possible
approximations, those based on expansions in powers of a small parameter
such as $\epsilon =d_{u}-d$ or $1/N$ (where the upper-dimension $d_{u}=4$
for the $N$-vector model) are perturbative in essence. They, however, present the
advantage of allowing analytic calculations but are attached to the
smallness of quantities that are actually not small in the cases of physical
interest. In some cases the perturbative framework may fail \cite{5745,5865}.

The derivative expansion \cite{212}, of present interest here, is
an expansion in powers of the derivative of the field. It is not associated
to a small parameter though it is expected to be rather adapted to the study
of phenomena at small momenta \footnote{%
A recent interesting attempt to adapt the derivative expansion to phenomena
that include effects at larger momenta is made in reference \cite{5872}.}
(large distances) like critical phenomena for instance. The interest of the
derivative expansion is that the physical parameters (like $d$ and $N$) may
take on arbitrary values. Hence, in the range of validity of the expansion
(thus presumed to be in the large distance regime), the approach is actually
nonperturbative. The drawback is the necessary recourse to numerical
techniques [for studying coupled nonlinear ordinary differential equations
(ODE)] that are not always well controlled. Consequently very few orders of
the derivative expansion have really been explicitly considered.

Many studies have been effectuated in the local potential approximation
(LPA), i.e. at the leading order [$O\left( \partial ^{0}\right) $] of the
derivative expansion \cite{4595,4700,4627}. Also, estimates of the critical
exponents for Ising-like models ($N=1$) have been obtained several times
from full studies of the next-to-leading order \footnote{%
In incomplete studies some contributions to a given order of the derivative
expansion are neglected. For example in reference \cite{3642}, despite an
estimation of the critical exponent $\eta $ (which is exactly equal to zero
at order $\partial ^{0}$), the order $\partial ^{2}$ has not been completely
considered because the evolution equation of the wave function
renormalization has been neglected. In reference \cite{4873}, $\mathrm{O}%
\left( N\right) $ systems are incompletely studied up to $O\left( \partial
^{2}\right) $ because one differential equation has been discarded. Also,
the study of reference \cite{5471} at order $\partial ^{4},$ though
interesting, is incomplete since only three differential equations, among
five constituting actually the order $\partial ^{4}$, have been
treated.} [i.e. $O\left( \partial
^{2}\right) $] \cite{212,3491,3357,4858,5633,5744}, and even from a full
study of the third order [$O\left( \partial ^{4}\right) $] \cite{5469}. On
the contrary, only two full studies have been effectuated up to $O\left( \partial
^{2}\right) $ for the $\mathrm{O}\left( N\right) $ vector model \cite%
{3828,4754}. Yet, this model provides us with the opportunity of
varying $N$ \ and of comparing the results with the best estimates of
critical exponents obtained from six \cite{3923,4006} or\ seven \cite{4228}
orders of the perturbative framework in a wide range of values of $N$.
Interesting informations on the possible convergence of the derivative
expansion are then reachable.

In the present study, I consider the $\mathrm{O}\left( N\right) $
Wilson-Polchinski exact RG equation expanded up to order $O\left( \partial
^{2}\right) $ in the derivative expansion and I calculate the critical
exponents. Beyond the specific calculation (which was lacking) the real aim
is to try to clarify (and also to evaluate) the present status of the
derivative expansion. Still, I must explain why I consider the
Wilson-Polchinski equation.

Indeed, there are several different approaches and treatments of the exact RG
equation so that it is not easy to really estimate the validity of the
choice of the equation or/and of the calculations available in the
litterature. Let me try to briefly summarize the situation and to justify my
choice.

There are two families of exact RG equations (for a review see \cite{4595}).
The first family expresses the RG flow ($\Lambda \mathrm{d}S_{\Lambda }/%
\mathrm{d}\Lambda $) of the microscopic action $S_{\Lambda }\left[  \phi %
\right] $ (hamiltonian) associated to a running momentum scale $\Lambda $
which, in the circumstances, is a running ultra-violet cutoff. The second
family expresses the RG flow of $\Gamma _{\Lambda }\left[ M\right] $, the
Legendre transform of $S_{\Lambda }\left[  \phi \right] $, in that case
the running scale $\Lambda $ effectively appears as an infra-red momentum
cutoff.

There is no fundamental difference between the two families since the object
of the RG is the same in the two cases: accounting for all the correlated scales $%
\Lambda $ ranging from 0 to $\infty $ (at criticality); only the physical meaning of the
field variable at hand has been changed: $ \phi $ is related to a
microscopic description (like a spin of the Ising model) while $M$ is
thought of as a macroscopic variable (like the magnetization). If one wants
to calculate an equation of state or some correlation functions or some
universal critical amplitude ratios, the second family of equations is
better adapted. But if one only wants to estimate critical exponents (for
example to illustrate the convergence of the derivative expansion), then
considering the first family is surely more efficient. Actually, the set of
ODE generated in the derivative expansion is much simpler when considered
with the first family than with the second \footnote{%
In the study of reference \cite{5469} of the Ising case up to $O\left(
\partial ^{4}\right) $, the consideration of an exact RG equation of the
second family yields a set of ODE the writing of
which requires 20 pages \cite{5804} while, even at order $\partial ^{6}$, the
first family yields a set of equations that holds on a half of page \cite%
{progress}. The complexity of dealing with the second family is also
well illustrated in the appendix of reference \cite{3828}.}. 
The first family is indeed better adapted to the calculation
of critical exponents for the same reason as in the field theoretical
approach to critical phenomena \cite{5459}, the critical exponents are
defined from renormalization functions that are introduced within the
microscopic action $S_{\Lambda }\left[  \phi \right] $. The Wilson \cite%
{440} and Polchinski \cite{354} exact RG equations belong to the first
family, they only differ by the way the smooth cutoff function has been
defined (a specific function for Wilson, an arbitrary one for Polchinski). Of
course, the two equations are physically equivalent, but due to a
misunderstanding in the introduction of the critical exponent $\eta $ in the
exact RG equation as formulated by Polchinski, it is only recently that the
equivalence has been clearly established \cite{3912,5744}.

If the coupled set of ODE generated by the derivative expansion is different
in the two families of exact RG equations, the treatments of the
differential equations encountered in the litterature differ also. For
convenience, let me classify the studies in two groups according to whether the authors
have adopted  
the conventional approach (defined below) or not.

The conventional approach is characterized as
follows:

\begin{enumerate}
\item the set of ODE is numerically studied as such (e.g., without
considering any artefact such as an expansion in powers of the field).

\item the critical exponent $\eta $ is introduced in a conventional way as
defined in references \cite{440,2727,5744}.

\item the critical exponents are estimated from a set of eigenvalue
equations linearized about a fixed point solution of the flow equation.

\item the reparametrization invariance is explicitly accounted for.
\end{enumerate}

To limit myself to the studies mentioned above that consider $\mathrm{O%
}\left( N\right) $ systems via equations of the second family developped up to
$O(\partial^2)$ \cite%
{3828,4754}: the study of reference \cite{3828} follows
the conventional approach while that of reference \cite{4754} does not. In
fact, in this latter work, except the first point above, none of the other
points is satisfied. This is particularly important relatively to point (iv)
because the reparametrization invariance induces a line of equivalent fixed
points along with $\eta $ is constant \cite{4420}. In the case where the invariance is
broken (it is generally the case within the derivative expansion, except in
reference \cite{3828}) then the fixed points along the
line are no longer equivalents and the effective $\eta $ (when
introduced conventionally) varies with the global normalisation of the field 
$ \phi $. Nevertheless, even if the invariance is broken, one expects that
a vestige of this invariance can still be observed \cite{4420,4421,212} via an extremum
of $\eta $ on varying the global normalization of the field $\phi$. 
The absence of explicit consideration of the reparametrization
invariance in reference \cite{4754} is intringuing inasmuch as the
estimation of the critical exponents is excellent (see section \ref{Comp}).

By considering the Wilson-Polchinski RG equation for $O(N)$ systems, my first aim is to
illustrate the conventional treatment as described above. Additionnal aims follow:

\begin{enumerate}
\item Morris and Turner \cite{3828} \ have imposed the reparametrization
invariance by choosing a specific cutoff function and the resulting
estimates of the critical exponents are not very good, especially for $%
\omega $. Does the Wilson-Polchinski RG equation also produce such bad
results at order $O\left( \partial ^{2}\right) $?.

\item Is the Wilson-Polchinski RG equation able to produce estimates of
critical exponents comparable to those obtained by Gersdorff and Wetterich 
\cite{4754}?

\item The $\mathrm{O}\left( N\right) $ exact RG equation expanded up to
order $\partial ^{2}$ involves three coupled ODE. Compared to the Ising case $%
N=1$, it is somewhat an intermediate state between the order $\partial ^{2}$
which involves two coupled equations and the order $\partial ^{4}$ which
involves five coupled equations \cite{5469,5471}.

\item Does one observe some signs of convergence of the derivative expansion already
at order $\partial^2$? 
\end{enumerate}

Relative to this latter point, it is worth mentioning here the work of Litim
 \cite{4714} whose aim, though very interesting, differs from that of the present work.
Litim focuses its attention almost exclusively on the second family of the exact RG equations 
and especially on the arbitrariness introduced by the regularization process (cut-off procedure)
but he does not account for the reparametrization
invariance.
 From general arguments (independent from the derivative expansion), he provides us with a criterium for choosing 
 the regulator which should optimize the convergence of the derivative expansion already at a very low order.
The considerations are surely useful especially for studying sophisticated systems, such as
gauge field theory for example, for which already the leading order of the derivative expansion 
is difficult to implement. However, Litim has not actually studied the convergence properties
of the derivative expansion in itself but, in fact, has implicitly assumed that it
converges (at least, the fact that the expansion could yield only asymptotic series has been excluded).
Furthermore, Litim's criterium of choice
does not apply to the Wilson-Polchinski RG equation. In particular, at leading order of the
derivative expansion, his optimization provides the same
estimates of critical exponents as those obtained with the Wilson-Polchinski RG equation (see section \ref{NumRes}) which, at this order,
does not display any dependence on the regularization process. 

\section{Derivative expansion up to O$\left( \partial ^{2}\right) $}

\subsection{Flow equations}

According to reference \cite{5744}, after subtraction of the high
temperature fixed point $\frac{1}{2}\sum_{a=1}^{N}\int_{q}\left( \tilde{P}%
\psi ^{2}\right) ^{-1}\phi _{q}^{a}\phi _{-q}^{a}$ from the action, the
Wilson-Polchinski exact RG equation satisfied by an $\mathrm{O}\left(
N\right) $-symmetric action $S\left[  \phi ^{a}\right] $ (with $a=1,\cdots
,N$) reads as follows: 
\begin{eqnarray}
\dot{S} &=&\sum_{a=1}^{N}\left\{ -\int_{q}\phi _{q}^{a}\left( \tilde{d}%
_{\phi }\,+2q^{2}\frac{\psi ^{\prime }}{\psi }+\mathbf{q}\cdot \partial
_{q}\right) \frac{\delta S}{\delta \phi _{q}^{a}}\right. \nonumber \\
&&\left. +\int_{q}\left( \varpi \tilde{P}-q^{2}\tilde{P}^{\prime }\right)
\psi ^{2}\left[ \frac{\delta ^{2}S}{\delta \phi _{q}^{a}\delta \phi _{-q}^{a}%
}-\frac{\delta S}{\delta \phi _{q}^{a}}\frac{\delta S}{\delta \phi _{-q}^{a}}%
\right] \right\}  \label{eq:WilPol00}
\end{eqnarray}%
in which $\dot{S}$ stands for $\mathrm{d}S/\mathrm{d}t=-\Lambda \mathrm{d}S/%
\mathrm{d}\Lambda $ (hence $\exp \left( -t\right) $ $=\Lambda /\Lambda _{0}$
in which $\Lambda _{0}$ is some initial momentum scale of reference \cite{5744}), $%
\mathbf{q}$ is a dimensionless $d$-vector ($d$ is the dimension of the
euclidean space and $\mathbf{q}=\left\{ q_{i},i=1,\cdots ,d\right\} $) , $%
q^{2}=\sum_{i=1}^{d}q_{i}^{2}$, $\tilde{P}(q^{2})$ is a dimensionless cutoff
function that decreases rapidly when $q\rightarrow \infty$ with $\tilde{P}(0)=1$ , $\psi \left(
q^{2}\right) $ is an arbitrary function (except the normalization $\psi
\left( 0\right) =1$) introduced to test the reparametrization invariance, $%
\varpi =1-\eta /2$ and $\tilde{d}_{\phi }=d/2+\varpi $. A prime
denotes a derivative with respect to $q^{2}$: $\psi ^{\prime }=\mathrm{d}%
\psi /\mathrm{d}q^{2}$, $\tilde{P}^{\prime }=\mathrm{d}\tilde{P}/\mathrm{d}%
q^{2}$
and $\mathbf{q}\cdot \partial _{q}f\left( q\right)
=\sum_{i=1}^{d}q_{i}\partial f/\partial q_{i}$.

The expansion up to order $\partial ^{2}$ consists in projecting equation (%
\ref{eq:WilPol00}) onto actions of the form: 
\begin{equation*}
{S[\phi ]}={\int d^{d}x\,}\left[ V{(\rho ,t)+}Z{(\rho ,t)(\partial \vec{\phi}%
)^{2}+}Y{(\rho ,t)(\vec{\phi}\partial \vec{\phi})^{2}}\right]
\end{equation*}%
with 
\begin{equation*}
\rho =\frac{1}{2}\sum_{\alpha =1}^{N}\phi ^{\alpha }\phi ^{\alpha }=\frac{1}{%
2}\vec{\phi}^{2}
\end{equation*}

Then the flow equations for $V$, $Z$ and $Y$ read as follows:%
\begin{eqnarray}
\dot{V} &=&I_{0}\left( N{V}^{\prime }{+2\rho V}^{\prime \prime }\right)
+dV-\left( d+2\varpi \right) \rho V^{\prime }-2\varpi \rho {V}^{\prime
2}  \nonumber \\
&&+2I_{1}\left( {N}Z+2{\rho }Y\right)  \nonumber \\
\dot{Z} &=&I_{0}\left( NZ^{\prime }{+2\rho Z^{\prime \prime }+2}Y\right)
-2\left( \varpi +1\right) Z{-}\left( d+2\varpi \right) \rho Z^{\prime
}-2\psi _{0}^{\prime }V^{\prime }  \nonumber \\
&&-4\varpi \left( {V}^{\prime }Z+\rho V^{\prime }Z^{\prime }\right) -\left[
\left( \varpi -1\right) \tilde{P}_{0}^{\prime }+2\varpi \psi _{0}^{\prime }%
\right] V^{\prime 2}  \nonumber \\
\dot{Y} &=&I_{0}\left( NY^{\prime }{+2\rho }Y^{\prime \prime }{+4}Y{^{\prime
}}\right) -(d+2+4\varpi )Y{-}\left( d+2\varpi \right) \rho Y^{\prime }-2\psi
_{0}^{\prime }{V}^{\prime \prime }  \nonumber \\
&&-4\varpi \left( {V}^{\prime \prime }Z+2{V}^{\prime }Y+\rho {V}^{\prime
}Y^{\prime }+2{\rho {V}^{\prime \prime }}Y\right)  \nonumber \\
&&-2\left[ \left( \varpi -1\right) \tilde{P}_{0}^{\prime }+2\varpi \psi
_{0}^{\prime }\right] \left( {V}^{\prime \prime }{V}^{\prime }+\rho {V}%
^{\prime \prime 2}\right)  \label{eq:VZY}
\end{eqnarray}%
in which a prime acting on $V$, $Z$ or $Y$ denotes this time a derivative
with respect to $\rho $, while $\psi _{0}^{\prime }\equiv \psi ^{\prime }\left(
0\right) $, $\tilde{P}_{0}^{\prime }\equiv \tilde{P}^{\prime }\left(
0\right) $ and: 
\begin{equation}
I_{0}=\int_{q}\left( \varpi \tilde{P}-q^{2}\tilde{P}^{\prime }\right) \psi
^{2},\qquad I_{1}=\int_{q}q^{2}\left( \varpi \tilde{P}-q^{2}\tilde{P}%
^{\prime }\right) \psi ^{2}  \label{eq:Integ}
\end{equation}

It is convenient to perform the following changes:%
\begin{equation}
\rho =NI_{0}\,\bar{\rho},\qquad V=NI_{0}\,\bar{V},\qquad Z=\bar{Z},\qquad Y=%
\frac{1}{NI_{0}}\,\bar{Y}  \label{eq:FacteurN}
\end{equation}%
and then to consider the new set of functions:%
\begin{equation*}
\mathrm{v}_{1}=\frac{\mathrm{d}\bar{V}}{\mathrm{d}\bar{\rho}},\qquad \mathrm{%
v}_{2}=\bar{Z},\qquad \mathrm{v}_{3}=\bar{Y}
\end{equation*}

Using these new notations and restoring the writing $\bar{\rho}%
\longrightarrow \rho $, the set of equations (\ref{eq:VZY}) becomes:%
\begin{eqnarray}
\mathrm{\dot{v}}_{1} &=&\left( 1+\frac{{2}}{N}\right) \mathrm{v}_{1}^{\prime
}{+}\frac{{2}}{N}{\rho \mathrm{v}_{1}^{\prime \prime }}-e\mathrm{v}%
_{1}-\left( d+e\right) \rho \mathrm{v}_{1}^{\prime }-e\left( \mathrm{v}%
_{1}^{2}+2\rho \mathrm{v}_{1}\mathrm{v}_{1}^{\prime }\right)  \nonumber \\
&&+P_{1}\left[ \mathrm{v}_{2}^{\prime }+\frac{2}{N}\left( \mathrm{v}_{3}+{%
\rho }\mathrm{v}_{3}^{\prime }\right) \right]  \label{eq:3eqs1} \\
\mathrm{\dot{v}}_{2} &=&\mathrm{v}_{2}^{\prime }{+}\frac{{2}}{N}{\rho 
\mathrm{v}_{2}^{\prime \prime }+}\frac{{2}}{N}\mathrm{v}_{3}-\left(
e+2\right) \mathrm{v}_{2}{-}\left( d+e\right) \rho \mathrm{v}_{2}^{\prime
}+2u\mathrm{v}_{1}  \nonumber \\
&&-2e\left( \mathrm{v}_{1}\mathrm{v}_{2}+\rho \mathrm{v}_{1}\mathrm{v}%
_{2}^{\prime }\right) +P_{2}\mathrm{v}_{1}^{2}  \label{eq:3eqs2} \\
\mathrm{\dot{v}}_{3} &=&\left( 1+\frac{{4}}{N}\right) \mathrm{v}_{3}^{\prime
}{+}\frac{{2}}{N}{\rho }\mathrm{v}_{3}^{\prime \prime }-(d+2+2e)\mathrm{v}%
_{3}{-}\left( d+e\right) \rho \mathrm{v}_{3}^{\prime }+2u\mathrm{v}%
_{1}^{\prime }  \nonumber \\
&&-2e\left[ \mathrm{v}_{1}^{\prime }\left( \mathrm{v}_{2}+2{\rho }\mathrm{v}%
_{3}\right) +\mathrm{v}_{1}\left( 2\mathrm{v}_{3}+\rho \mathrm{v}%
_{3}^{\prime }\right) \right] +2P_{2}\left( \mathrm{v}_{1}+\rho \mathrm{v}%
_{1}^{\prime }\right) \mathrm{v}_{1}^{\prime }  \label{eq:3eqs3}
\end{eqnarray}%
in which:%
\begin{eqnarray*}
u &=&-\psi _{0}^{\prime },\qquad e=2\varpi \qquad  \\
P_{1} &=&2\frac{I_{1}}{I_{0}},\qquad P_{2}=-\left( \frac{e}{2}-1\right) 
\tilde{P}^{\prime }(0)+eu
\end{eqnarray*}

\subsection{Fixed point equations}

The fixed point equations correspond to the three simultaneous conditions 
$\mathrm{\dot{v}}_{i}=0$ for $i=1,2,3$ which yield three coupled nonlinear
ODE of second order each:

\begin{eqnarray}
{\mathrm{v}_{1}^{\prime \prime }} &=&\frac{N}{2\rho }\left[ e\mathrm{v}%
_{1}\left( 1+\mathrm{v}_{1}\right) -\left( 1+\frac{{2}}{N}\right) \mathrm{v}%
_{1}^{\prime }-P_{1}\left( \mathrm{v}_{2}^{\prime }+\frac{2}{N}\mathrm{v}%
_{3}\right) \right]  \nonumber \\
&&+\frac{N}{2}\left( d+e+2e\mathrm{v}_{1}\right) \mathrm{v}_{1}^{\prime
}-P_{1}\mathrm{v}_{3}^{\prime }  \label{eq:3eqsPF1} \\
{\mathrm{v}_{2}^{\prime \prime }} &=&\frac{N}{2\rho }\left[ \left( e+2+2e%
\mathrm{v}_{1}\right) \mathrm{v}_{2}-\mathrm{v}_{2}^{\prime }{-}\frac{{2}}{N}%
\mathrm{v}_{3}-2u\mathrm{v}_{1}-P_{2}\mathrm{v}_{1}^{2}\right]  \nonumber \\
&&+\frac{N}{2}\left[ \left( d+e\right) {+}2e\mathrm{v}_{1}\right] \mathrm{v}%
_{2}^{\prime }  \label{eq:3eqsPF2} \\
\mathrm{v}_{3}^{\prime \prime } &=&\frac{N}{2\rho }\left[ \left( d+2+2e+4e%
\mathrm{v}_{1}\right) \mathrm{v}_{3}-\left( 1+\frac{{4}}{N}\right) \mathrm{v}%
_{3}^{\prime }+2\left( e\mathrm{v}_{2}-u-P_{2}\mathrm{v}_{1}\right) \mathrm{v%
}_{1}^{\prime }\right]  \nonumber \\
&&{+}\frac{N}{2}\left[ \left( d+e+2e\mathrm{v}_{1}\right) \mathrm{v}%
_{3}^{\prime }+2\left( 2e\mathrm{v}_{3}-P_{2}\mathrm{v}_{1}^{\prime }\right) 
\mathrm{v}_{1}^{\prime }\right]  \label{eq:3eqsPF3}
\end{eqnarray}

The differential system is of order six, thus the general solution depends
on six arbitrary constants. Three of these constants are fixed so as to
avoid the singularity at the origin $\rho =0$ displayed by the equations,
hence the three following conditions:

\begin{eqnarray}
\mathrm{v}_{2}^{\prime }\left( 0\right)  &=&\left[ e+2+2e\mathrm{v}%
_{1}\left( 0\right) \right] \mathrm{v}_{2}\left( 0\right) {-}\frac{{2}}{N}%
\mathrm{v}_{3}\left( 0\right) -\mathrm{v}_{1}\left( 0\right) \left[ 2u+P_{2}%
\mathrm{v}_{1}\left( 0\right) \right]   \label{eq:OriCond1} \\
\mathrm{v}_{1}^{\prime }\left( 0\right)  &=&\frac{N}{N+2}\left\{ e\mathrm{v}%
_{1}\left( 0\right) \left[ 1+\mathrm{v}_{1}\left( 0\right) \right] -P_{1}%
\left[ \mathrm{v}_{2}^{\prime }\left( 0\right) +\frac{2}{N}\mathrm{v}%
_{3}\left( 0\right) \right] \right\}   \label{eq:OriCond2} \\
\mathrm{v}_{3}^{\prime }\left( 0\right)  &=&\frac{N}{N+4}\left\{ \left[
d+2+2e+4e\mathrm{v}_{1}\left( 0\right) \right] \mathrm{v}_{3}\left( 0\right)
\right.   \nonumber \\
&&\left. +2\left[ e\mathrm{v}_{2}\left( 0\right) -u-P_{2}\mathrm{v}%
_{1}\left( 0\right) \right] \mathrm{v}_{1}^{\prime }\left( 0\right) \right\} 
\label{eq:OriCond3}
\end{eqnarray}

If $\eta $ is a priori fixed, then the general solution of
the set of equations (\ref{eq:3eqsPF1}--\ref{eq:3eqsPF3}) depends on the
three remaining arbitrary constants, e.g. the values $\mathrm{v}_{i}\left(
0\right) $ for $i=1,2,3$. In general the corresponding solutions are
singular at some varying $\rho ^{\ast }$ (moving singularity), with:%
\begin{eqnarray*}
\mathrm{v}_{1}\left( \rho \right) \propto \left( \rho ^{\ast }-\rho
\right) ^{-1},\qquad 
\mathrm{v}_{2}\left( \rho \right) \propto \left( \rho ^{\ast }-\rho
\right) ^{-2},\qquad 
\mathrm{v}_{3}\left( \rho \right) \propto \left( \rho ^{\ast }-\rho
\right) ^{-2}
\end{eqnarray*}

But, the equations (\ref{eq:3eqsPF1}--\ref{eq:3eqsPF3}) admit another kind of solution that goes to infinity ($\rho
\rightarrow \infty $) without encountering any singularity and which behaves
asymptotically for large $\rho $ as follows:%
\begin{eqnarray}
\mathrm{v}_{1\mathrm{asy}}\left( \rho \right)  &=&G_{1}\rho ^{\theta
_{1}}+\left( 1+2\theta _{1}\right) G_{1}^{2}\rho ^{2\theta _{1}}+\cdots 
\label{eq:v1asy} \\
\mathrm{v}_{2\mathrm{asy}}\left( \rho \right)  &=&uG_{1}\rho ^{\theta
_{1}}+G_{1}^{2}\,\left( 1+{{\theta }_{1}}\right) \frac{2\,\left(
d\,-\,e\right) \,u+\left( d\,+e\,\right) {\tilde{P}}_{0}^{\prime }}{2\,d}%
\rho ^{2\theta _{1}} \nonumber \\
&&+G_{2}\rho ^{\theta _{2}}+\cdots   \label{eq:v2asy} \\
\mathrm{v}_{3\mathrm{asy}}\left( \rho \right)  &=&u\theta _{1}G_{1}\rho
^{\theta _{1}-1}+2G_{1}^{2}\,{{\theta }_{1}}\,\frac{2\,\left( d\,-\,e\right)
\,u+d\,{\tilde{P}}^{\prime }{_{0}}}{2\,\left( d+e\right) }\rho ^{2\theta
_{1}-1}\nonumber \\
&&+G_{3}\rho ^{\theta _{3}}+\cdots   \label{eq:v3asy}
\end{eqnarray}%
with:%
\begin{eqnarray*}
\theta _{1} =-\frac{e}{d+e},\qquad 
\theta _{2} =-\frac{e+2}{d+e},\qquad 
\theta _{3} =\theta _{2}-1
\end{eqnarray*}

The values of the three constants $\left\{ G_{i};i=1,2,3\right\} $,
correspond to some adjustment of the set $\left\{ \mathrm{v}%
_{i}\left( 0\right) ;i=1,2,3\right\} $ and vice versa. This nonsingular
solution is the fixed point solution which we are interested in. When $\eta $
is a priori fixed, the six arbitrary constants of integration are then
determined, the differential system is balanced.

If $\eta $ is considered as an unknown parameter to be determined, then one
of the three preceding quantities $\left\{ \mathrm{v}_{i}\left( 0\right)
\right\} $ or $\left\{ G_{i}\right\} $ must be promoted to
the rank of a fixed parameter chosen a priori. In general one chooses 
\begin{equation}
\mathrm{v}_{2}\left( 0\right) =Z_{0}  \label{eq:Z0}
\end{equation}%
which corresponds to having fixed to $Z_{0}$ the value of the kinetic term
in $S\left[  \phi ^{a}\right] $ and is customarily associated with the
arbitrary global normalization of the field $\phi$. One thus obtains a function $\eta \left(
Z_{0}\right) $ which should be a constant if the reparametrization
invariance of the exact RG equation was preserved by the derivative
expansion presently considered ($\eta $ should be a constant along a line of
equivalent fixed points generated by the variation of $Z_{0}$). Since it is
not the case, one actually obtains a nontrivial function $\eta \left(
Z_{0}\right) $. Fortunately a vestige of the reparametrisation invariance is
preserved and $\eta \left( Z_{0}\right) $ displays an extremum in $Z_{0}$.
This provides us with an optimal value ($\eta ^{\mathrm{opt}}$) of $\eta $
(and similarly for $Z_{0}$) via the condition:%
\begin{equation}
\frac{\mathrm{d}\eta ^{\mathrm{opt}}}{\mathrm{d}Z_{0}}=0  \label{eq:EtaOpt}
\end{equation}

Instead of using this condition to determine $\eta ^{\mathrm{opt}}$, I use
the fact that the line of equivalent fixed points is associated with a
redundant operator with a zero eigenvalue\footnote{The fact that the eigenvalue takes on a definite value although it is associated
with a redundant operator is not in conflict with the work of Wegner (see reference \cite{2835} and references therein)
which indicates that the eigenvalue of a redundant operator generally varies with the renormalization process.
As shown in reference \cite{4420},
the linear character of the renormalization of the field in the process of generating the exact RG equation implies
a definite eigenvalue (the 
reparametrization invariance is a direct consequence of this linearity). On the contrary, in the case
of a nonlinear renormalization scheme, the eigenvalue in question no longer is constant but depends
on the renormalization procedure \cite{4420} in accordance with \cite{2835}.} \cite{4420,4421}. 
Hence, one may determine $\eta ^{%
\mathrm{opt}}$ by imposing that the fixed point of interest be associated
to a zero eigenvalue. This leads us to the consideration of the system of
eigenvalue equations.

\subsection{Eigenvalue equations}

The eigenvalue equations are obtained by linearization of the flow equations
(\ref{eq:3eqs1}--\ref{eq:3eqs3}) about a fixed point solution $\left\{ 
\mathrm{v}_{i}^{\mathrm{\ast }},i=1,2,3\right\} $:%
\begin{equation*}
\mathrm{v}_{i}=\mathrm{v}_{i}^{\mathrm{\ast }}+\varepsilon \mathrm{e}%
^{\lambda t}\mathrm{g}_{i}
\end{equation*}

Keeping the linear contribution in $\varepsilon $, the following set of
coupled ODE comes:%
\begin{eqnarray}
\mathrm{g}_{1}^{\prime \prime } &=&\frac{N}{2\rho }\left[ \left( \lambda
+e+2e\mathrm{v}_{1}^{\mathrm{\ast }}\right) \mathrm{g}_{1}-\left( 1+\frac{{2}%
}{N}\right) \mathrm{g}_{1}^{\prime }-P_{1}\left( \mathrm{g}_{2}^{\prime }+%
\frac{2}{N}\mathrm{g}_{3}\right) \right]   \nonumber \\
&&+\frac{N}{2}\left[ \left( d+e+2e\mathrm{v}_{1}^{\mathrm{\ast }}\right) 
\mathrm{g}_{1}^{\prime }+2e\mathrm{g}_{1}\mathrm{v}_{1}^{\mathrm{\ast }%
\prime }\right] -P_{1}\mathrm{g}_{3}^{\prime }  \label{eq:VP01} \\
{\mathrm{g}_{2}^{\prime \prime }} &=&\frac{N}{2\rho }\left[ \left( \lambda
+e+2+2e\mathrm{v}_{1}^{\mathrm{\ast }}\right) \mathrm{g}_{2}-\mathrm{g}%
_{2}^{\prime }{-}\frac{{2}}{N}\mathrm{g}_{3}+2\left( e\mathrm{v}_{2}^{\ast
}-u-P_{2}\mathrm{v}_{1}^{\mathrm{\ast }}\right) \mathrm{g}_{1}\right]  
\nonumber \\
&&{+}\frac{N}{2}\left[ \left( d+e+2e\mathrm{v}_{1}^{\mathrm{\ast }}\right) 
\mathrm{g}_{2}^{\prime }+2e\mathrm{g}_{1}\mathrm{v}_{2}^{\ast \prime }\right]
\label{eq:VP02} \\
\mathrm{g}_{3}^{\prime \prime } &=&\frac{N}{2\rho }\left[ (\lambda +d+2+2e+4e%
\mathrm{v}_{1}^{\ast })\mathrm{g}_{3}-\left( 1+\frac{{4}}{N}\right) \mathrm{g%
}_{3}^{\prime }-2u\mathrm{g}_{1}^{\prime }\right.   \nonumber \\
&&\left. +2e\left( \mathrm{g}_{1}^{\prime }\mathrm{v}_{2}^{\ast }+2\mathrm{g}%
_{1}\mathrm{v}_{3}^{\ast }\right) -2P_{2}\left( \mathrm{g}_{1}\mathrm{v}%
_{1}^{\ast \prime }+\mathrm{v}_{1}^{\ast }\mathrm{g}_{1}^{\prime }\right) +2e%
\mathrm{v}_{1}^{\ast \prime }\mathrm{g}_{2}\right]   \nonumber \\
&&{+}\frac{N}{2}\left[ \left( d+e+2e\mathrm{v}_{1}^{\mathrm{\ast }}\right) 
\mathrm{g}_{3}^{\prime }+2e\left( 2\mathrm{g}_{1}^{\prime }\mathrm{v}%
_{3}^{\ast }+\mathrm{g}_{1}\mathrm{v}_{3}^{\ast \prime }+2\mathrm{v}%
_{1}^{\ast \prime }\mathrm{g}_{3}\right) -4P_{2}\mathrm{g}_{1}^{\prime }%
\mathrm{v}_{1}^{\ast \prime }\right]   \label{eq:VP03}
\end{eqnarray}

Similar considerations as those relative to the determination of the six
integration constants associated to the fixed point equations (\ref%
{eq:3eqsPF1}--\ref{eq:3eqsPF3}) stand. For a given fixed point solution ($\mathrm{v}_i^*$),
there remain six constants to be determined. Three of them are fixed so as
to avoid the singularity at the origin $\rho =0$ displayed by the equations:%
\begin{eqnarray*}
\mathrm{g}_{2}^{\prime }\left( 0\right) &=&\left[ \lambda +e+2+2e\mathrm{v}%
_{1}^{\mathrm{\ast }}\left( 0\right) \right] \mathrm{g}_{2}\left( 0\right) {-%
}\frac{{2}}{N}\mathrm{g}_{3}\left( 0\right) \\
&&+2\left[ e\mathrm{v}_{2}^{\ast }\left( 0\right) -u-P_{2}\mathrm{v}_{1}^{%
\mathrm{\ast }}\left( 0\right) \right] \mathrm{g}_{1}\left( 0\right) \\
\mathrm{g}_{1}^{\prime }\left( 0\right) &=&\frac{N}{N+2}\left\{ \left[
\lambda +e+2e\mathrm{v}_{1}^{\mathrm{\ast }}\left( 0\right) \right] \mathrm{g%
}_{1}\left( 0\right) -P_{1}\left[ \mathrm{g}_{2}^{\prime }\left( 0\right) +%
\frac{2}{N}\mathrm{g}_{3}\left( 0\right) \right] \right\} \\
\mathrm{g}_{3}^{\prime }\left( 0\right) &=&\frac{N}{N+4}\left\{ \left[
\lambda +d+2+2e+4e\mathrm{v}_{1}^{\ast }\left( 0\right) \right] \mathrm{g}%
_{3}\left( 0\right) -2u\mathrm{g}_{1}^{\prime }\left( 0\right) \right. \\
&&+2e\left[ \mathrm{g}_{1}^{\prime }\left( 0\right) \mathrm{v}_{2}^{\ast
}\left( 0\right) +2\mathrm{g}_{1}\left( 0\right) \mathrm{v}_{3}^{\ast
}\left( 0\right) \right] \nonumber \\
&&\left.-2P_{2}\left[ \mathrm{g}_{1}\left( 0\right) \mathrm{%
v}_{1}^{\ast \prime }\left( 0\right) +\mathrm{v}_{1}^{\ast }\left( 0\right) 
\mathrm{g}_{1}^{\prime }\left( 0\right) \right] 
 +2e\mathrm{v}_{1}^{\ast \prime }\left( 0\right) \mathrm{g}%
_{2}\left( 0\right) \right\}
\end{eqnarray*}

One is interested in the solution that is regular when $\rho \rightarrow
\infty $. 

For $\lambda $ a priori fixed, the three values $\left\{ \mathrm{g}%
_{i}\left( 0\right) ,i=1,2,3\right\} $ at the origin $\rho =0$ must be
adjusted so that the solution reaches the following regular asymptotic behavior:

\begin{eqnarray*}
\mathrm{g}_{1\mathrm{asy}} &=&S_{1}\rho ^{\chi _{1}}+2\,\,\left( 1+{{\theta }%
_{1}+}\chi _{1}\right) \,{G_{1}}\,{S_{1}\rho }^{{{\theta }_{1}+}\chi
_{1}}+\cdots \\
\mathrm{g}_{2\mathrm{asy}} &=&S_{2}\rho ^{\chi _{2}}+uS_{1}\rho ^{\chi
_{1}}+\cdots \\
\mathrm{g}_{3\mathrm{asy}} &=&S_{3}\rho ^{\chi _{3}}+uS_{1}\chi _{1}\rho
^{\chi _{1}-1}+\cdots
\end{eqnarray*}%
with:%
\begin{eqnarray*}
\chi _{1} =-\frac{\lambda +e}{d+e}, \qquad
\chi _{2} =-\frac{\lambda +e+2}{d+e}, \qquad
\chi _{3} =\chi _{2}-1
\end{eqnarray*}%
and the value of the set of constants $\left\{ S_{i},i=1,2,3\right\} $
entering the regular solution at large $\rho $ corresponds to the value of $\left\{\mathrm{g}_{i}(0)\right\} $
adjusted at the origin and vice versa.

As in any eigenvalue problem, the global normalization of the eigenvector
may be chosen at will so that, fixing $\mathrm{g}_{1}\left( 0\right) =1$ for
instance, allows one to determine discrete values of $\lambda $. Positive values
give the critical exponents, negative values are subcritical (or correction-to-scaling)
exponents.

The peculiar value $\lambda =0$, if present, is associated to the vestige of
the reparametrization invariance \cite{4421,212}. Indeed this zero eigenvalue is associated
to the redundant operator that generates the line of equivalent fixed points
in the complete theory \cite{4420,4421}. Conversely, if one considers together the fixed
point equations (\ref{eq:3eqsPF1}--\ref{eq:3eqsPF3}) with the eigenvalue
equations (\ref{eq:VP01}--\ref{eq:VP03}) in which $\lambda $ is set equal to
zero (and the condition $\mathrm{g}_{1}\left( 0\right) =1$ is maintained), then
the condition (\ref{eq:Z0}) may be abandoned and $\mathrm{v}_{2}\left(
0\right) $ adjusted so as to get a common solution to the set of six
coupled ODE. Then, the resulting value of $\eta $ nececessarily coincides
with $\eta ^{\mathrm{opt}}$ as defined by equation (\ref{eq:EtaOpt}) and the 
resulting value of $\mathrm{v}_2(0)$ gives $Z_0^{\mathrm{opt}}$. Though
the number of differential equations has increased twofold this procedure of
determining the optimized fixed point is the most efficient one when parameters
(like $N$ and some other ones, see following section) have to be varied.

\subsection{The free parameters}

In order to perform an actual numerical study of the set of second order ODE
described in the preceding section, I make the following choice \footnote{%
In reference \cite{5744}, I considered one additionnal parameter within the
cutoff function $\tilde{P}$ (named $a$), however the equations were
invariant in the change $b\rightarrow b/a$ so that one of the two parameters
was unnecessary.}: 
\begin{eqnarray}
\tilde{P}\left( q^{2}\right)  &=&\mathrm{e}^{-q^{2}} \label{eq:P}\\
\psi \left( q^{2}\right)  &=&\frac{1}{1+bq^{2}}
\end{eqnarray}

Following the terminology of reference \cite{2835}, the free parameter $b$
is redundant and is intended to be used to optimize the numerical results of
the derivative expansion.

The introduction of $b$ is linked to the general property of
reparametrization invariance which is broken by the present derivative
expansion. The normalization $\psi \left( 0\right) =1$ is chosen in order to
distinguish the effect of simply changing the global normalization of the
field which induces a line of equivalent fixed points (at fixed $b$). That
line is customarily associated to the arbitrariness of $Z_{0}$ the value of $%
Z\left( 0\right) $ ($\equiv \mathrm{v}_{2}\left( 0\right) $), i.e. the
coefficient of the kinetic term in $S\left[  \phi ^{a}\right] $ (see
preceding section). Changing the value of $b$ in the complete theory would
induce new (equivalent) lines of equivalent fixed points.

Though they are part of the same invariance, the two free parameters $b$ and 
$Z_{0}$ have effects of different nature in the derivative expansion. As a
global constant of normalization, one can expect that, at a given order of
the derivative expansion, the variation of $Z_{0}$ will still reveal a
vestige of the invariance of the exact theory (see preceding section). On
the contrary, the effect of $b$ spreads out over different orders of the
derivative expansion. Consequently, one expects to observe a progressive restoration
of the redundant character of $b$ as the order of the expansion increases.
 Regarding the extremely low order\ considered here ($%
\mathrm{O}\left( \partial ^{2}\right) $), one must not expect too much from
varying $b$ (see section \ref{NumRes}).

Notice that the cutoff function $\tilde{P}\left( q^{2}\right) $ is essentially
a regulator of the integrals (\ref{eq:Integ}) generated by the derivative expansion.
Besides the arbitrary choice in the decreasing at large $q$, the
other sources of arbitrariness of the cutoff function may be included within the arbitrary
function $\psi \left( q^{2}\right) $. This is why the choice (\ref{eq:P}) does not
involve any free parameter.

\section{Numerical study and results}

There are two different methods for numerically studying systems of coupled nonlinear
ODE as those described above: the shooting and the relaxation methods (see,
for example reference \cite{4394}). Because it is the easiest to implement,
only the shooting method is considered here (though it is less numerically
stable than the relaxation method).

\subsection{The shooting method}

Considering an initial point $\rho _{1}$ where known conditions (initial
conditions) are imposed and trial values are given to the remaining
integration constants, one integrates the ODE system up to a second point $%
\rho _{2}$ where the required conditions are checked. Using a Newton-Raphson
algorithm, one iterates the test until the latter conditions are satisfied
within a given accuracy.

In the present study, the two points $\rho _{1}$ and $\rho _{2}$ are either
the origin $\rho =0$ and a large value $\rho _{\mathrm{asy}}$ (shooting from
the origin) or the reverse (shooting to the origin).

In principle, shooting to the origin is technically better adapted to the
present study. For example, in the case of the fixed point equations (\ref%
{eq:3eqsPF1}--\ref{eq:3eqsPF3}) with $\eta $ fixed, one starts from $\rho _{%
\mathrm{asy}}$ with trial values for the three constants $\left\{
G_{i}\right\} $ and the initial values of the three functions and of their
first derivatives $\left\{ \mathrm{v}_{i},\mathrm{v}_{i}^{\prime }\right\} $
defined by (\ref{eq:v1asy}--\ref{eq:v3asy}). After integration up to the
origin, one checks whether the three conditions (\ref{eq:OriCond1}--\ref%
{eq:OriCond3}) are fulfilled or not. The system is balanced and the values
of the other interesting parameters are simply by-products of the
adjustment. For example, the value of $Z_{0}$ associated to the arbitrarily
fixed value of $\eta $ is simply read as the value that $\mathrm{v}%
_{2}\left( 0\right) $ takes on after achievement of the adjustment. If one
wants instead to determine $\eta \left( Z_{0}\right) $, then $\eta $ has to
be considered as a trial parameter like $\left\{ G_{i}\right\} $ and the
supplementary condition (\ref{eq:Z0}) with $Z_{0}$ an arbitrary fixed
number, must be fulfilled at the origin.

Unfortunately, equations (\ref{eq:3eqsPF1}--\ref{eq:3eqsPF3}) are singular
at $\rho =0$ so that it is impossible to shoot to the origin. In reference \cite{3860}
(where the leading order LPA was studied for small values of $N$), the
difficulty was circumvented by shooting to a point close to the origin. On
the contrary, I have chosen to shoot from the origin because it is easy to
control the equations starting from that point.

Starting from the origin implies that, for a given value of $\eta $, the
adjustable parameters no longer are the $\left\{ G_{i}\right\} $ but the
initial values $\left\{ \mathrm{v}_{i}\left( 0\right) \right\} $, the
initial values of the first derivatives follow from (\ref{eq:OriCond1}--\ref%
{eq:OriCond3}). At the point $\rho _{\mathrm{asy}}$ only three conditions
are needed in order to balance the number of adjustable parameters. Thus one must
eliminate the $\left\{ G_{i}\right\} $ from the six equations (\ref%
{eq:v1asy}--\ref{eq:v3asy}) to obtain three anonymous conditions. Consequently
the precise knowledge of the asymptotic behavior of the regular fixed point
solution is not necessary. 

It appears that imposing conditions like $\left\{ 
\mathrm{v}_{i}^{\prime \prime }\left( \rho _{\mathrm{asy}}\right) =0\right\} 
$ is sufficient to determine the solution of interest. I first adjust the
trial parameters so as to reach some not too large a value of $\rho _{%
\mathrm{asy}}$, then I increase $\rho _{\mathrm{asy}}$ until the desired
accuracy is reached on the trial parameters.

\subsection{Numerical results}
\label{NumRes}

\begin{table}
\caption{\label{Tab1} Critical exponent estimates (for $d=3$) from the $\mathrm{O}\left( N\right)$ 
Wilson-Polchinski exact RG equation expanded up to $O\left( \partial ^{2}\right) $ in the derivative expansion
  for two values of the free parameter $b$.}
\begin{indented}
\item[]\begin{tabular}{lllllll}
\br
& \multicolumn{3}{l}{$b=0.03$} & \multicolumn{3}{l}{$b=0.11$} \\ \hline
$N$ & $\eta $ & $\nu $ & $\omega $ & $\eta $ & $\nu $ & $\omega $ \\ \mr
\multicolumn{1}{l}{$1$} & $0.01006$ & $0.6223$ & $0.7755$ & $0.02494$ & $%
0.5994$ & $0.8740$ \\ 
\multicolumn{1}{l}{$2$} & $0.00866$ & $0.6723$ & $0.7266$ & $0.02263$ & $%
0.6388$ & $0.7656$ \\ 
\multicolumn{1}{l}{$3$} & $0.00721$ & $0.7238$ & $0.7132$ & $0.02007$ & $%
0.6838$ & $0.7156$ \\ 
\multicolumn{1}{l}{$4$} & $0.00592$ & $0.7713$ & $0.7223$ & $0.01739$ & $%
0.7313$ & $0.7010$ \\ 
\multicolumn{1}{l}{$5$} & $0.00486$ & $0.8111$ & $0.7437$ & $0.01489$ & $%
0.7762$ & $0.7088$ \\ 
\multicolumn{1}{l}{$6$} & $0.00404$ & $0.8424$ & $0.7695$ & $0.01275$ & $%
0.8138$ & $0.7307$ \\ 
\multicolumn{1}{l}{$7$} & $0.00342$ & $0.866$ & $0.7946$ & $0.01100$ & $%
0.8440$ & $0.7576$ \\ 
\multicolumn{1}{l}{$8$} & $0.00294$ & $0.8849$ & $0.8172$ & $0.00960$ & $%
0.8673$ & $0.7842$ \\ 
\multicolumn{1}{l}{$9$} & $0.00256$ & $0.8993$ & $0.8368$ & $0.00848$ & $%
0.8854$ & $0.8081$ \\ 
\multicolumn{1}{l}{$10$} & $0.00227$ & $0.9107$ & $0.8533$ & $0.00757$ & $%
0.8995$ & $0.8292$ \\ 
\multicolumn{1}{l}{$11$} & $0.00204$ & $0.9214$ & $0.8711$ & $0.00683$ & $%
0.9108$ & $0.8463$ \\ 
\multicolumn{1}{l}{$12$} & $0.00184$ & $0.9293$ & $0.8836$ & $0.00621$ & $%
0.9198$ & $0.8613$ \\ 
\multicolumn{1}{l}{$15$} & $0.00142$ & $0.9438$ & $0.9060$ & $0.00488$ & $%
0.9387$ & $0.8942$ \\ 
\multicolumn{1}{l}{$20$} & $0.00102$ & $0.9596$ & $0.9330$ & $0.00358$ & $%
0.9563$ & $0.9258$ \\ \br
\end{tabular}
\end{indented}
\end{table}

\begin{table}
\caption{\label{Tab2} Same as table \ref{Tab1} (two other values of $b$).}
\begin{indented}
\item[] \begin{tabular}{lllllll}
\br
& \multicolumn{3}{l}{$b=0.17011$} & \multicolumn{3}{l}{$b=0.25611$} \\ 
\hline
$N$ & $\eta $ & $\nu $ & $\omega $ & $\eta $ & $\nu $ & $\omega $ \\ \mr
\multicolumn{1}{l}{$1$} & $0.03553$ & $0.5850$ & $0.9515$ & $0.04979$ & $%
0.5677$ & $1.0906$ \\ 
\multicolumn{1}{l}{$2$} & $0.03278$ & $0.6176$ & $0.7917$ & $0.04670$ & $%
0.5908$ & $0.8273$ \\ 
\multicolumn{1}{l}{$3$} & $0.02980$ & $0.6574$ & $0.7115$ & $0.04371$ & $%
0.6242$ & $0.7288$ \\ 
\multicolumn{1}{l}{$4$} & $0.02643$ & $0.7033$ & $0.6806$ & $0.03999$ & $%
0.6664$ & $0.6570$ \\ 
\multicolumn{1}{l}{$5$} & $0.02300$ & $0.7506$ & $0.6809$ & $0.03565$ & $%
0.7175$ & $0.6538$ \\ 
\multicolumn{1}{l}{$6$} & $0.01986$ & $0.7939$ & $0.7040$ & $0.03175$ & $%
0.7737$ & $0.6682$ \\ 
\multicolumn{1}{l}{$7$} & $0.01719$ & $0.8291$ & $0.7331$ & $0.02772$ & $%
0.8224$ & $0.7137$ \\ 
\multicolumn{1}{l}{$8$} & $0.01502$ & $0.8565$ & $0.7641$ & $0.02352$ & $%
0.8565$ & $0.7456$ \\ 
\multicolumn{1}{l}{$9$} & $0.01327$ & $0.8779$ & $0.7938$ & $0.02150$ & $%
0.8928$ & $0.8037$ \\ 
\multicolumn{1}{l}{$10$} & $0.01185$ & $0.8944$ & $0.8194$ & $0.01937$ & $%
0.9132$ & $0.8165$ \\ 
\multicolumn{1}{l}{$11$} & $0.01068$ & $0.9075$ & $0.8413$ & $0.01734$ & $%
0.9357$ & $0.8505$ \\ 
\multicolumn{1}{l}{$12$} & $0.00972$ & $0.9180$ & $0.8596$ & $0.01588$ & $%
0.9552$ & $0.8741$ \\ 
\multicolumn{1}{l}{$15$} & $0.00779$ & $0.9552$ & $0.9209$ & --- & --- & 
--- \\ \br
\end{tabular}
\end{indented}
\end{table}

\begin{table}
\caption{\label{Tab3} Compared to LPA on the lhs of the arrows, the order $\partial ^{2}$ (rhs of the arrows) induces a splitting into two
 of the subcritical exponents of degree higher than $\omega$
(values for $d=3$, $N=1$ 
  and $b=0.11$). The LPA values may be found in \cite{5252}.}
\begin{indented}
\item[]\begin{tabular}{llll}
\br
$\omega $ & $\omega _{2}$ & $\omega _{3}$ & $\omega _{4}$ \\ \mr
$0.6557\rightarrow 0.8740$ & $3.18\rightarrow \left\{ 
\begin{array}{l}
2.83 \\ 
3.62%
\end{array}%
\right. $ & $5.91\rightarrow \left\{ 
\begin{array}{l}
5.70 \\ 
6.40%
\end{array}%
\right. $ & $8.80\rightarrow \left\{ 
\begin{array}{l}
7.30 \\ 
9.6%
\end{array}%
\right. $ \\ \br
\end{tabular}
\end{indented}
\end{table}

At the leading order LPA, the function $\mathrm{v}_{1}$ is considered alone.
Only one second order ODE defines the fixed point (twice this number for the
eigenvalue problem). The study has already been done in reference \cite{3860}
for $d=3$ and $N=1$ to $4$. I have extended the results (for $d=3$) to larger values of $N$. The study is
simple because:
\begin{enumerate}
 \item the reparametrization invariance is automatically
satisfied since, by construction, the coefficient of the kinetic term is supposed to be constant.
\item the equation (\ref{eq:3eqs1}) with $\mathrm{v}_{2}=\mathrm{v}_{3}=0$, does not depend on any free parameter like $b$.
\end{enumerate} 

Consequently there is no ambiguity: to any value of $N$, corresponds a unique value of
$\nu$ and $\omega$ while $\eta$=0. As already mentioned in reference \cite{5049}, 
it is noteworthy that the
values I obtain (see also \cite{3860}) for the critical and subcritical exponents (without
optimization since there is no free parameters at hand) agree,
\textquotedblleft to all published digits\textquotedblright , with the
\textquotedblleft optimized\textquotedblright\ values obtained in reference 
\cite{5252} from a study of an exact RG equation of the second family. 
Hence, for the numerical results I have obtained at order LPA for $\nu$, $\omega$ and other subleading critical exponents $\omega_n$,
the reader is referred to table 1 of reference \cite{5252}.

At order $\partial ^{2}$, once the arbitrariness of $Z_0$ has been removed via the definition of $\eta ^{\mathrm{opt}}$,
  the values of the critical exponents still depend on $b$
in such a way that, for a given value of $N$, it is impossible to define \textquotedblleft
preferred\textquotedblright\ values. For example $\eta$ (from now on, $\eta$ stands for $\eta ^{\mathrm{opt}}$) depends
on $b$ almost linearly. 

Due to the spreading out over several orders of the effects of
$b$, the study of the convergence of the derivative expansion
relies rather on the consideration of several orders. Now the order $%
\partial ^{2}$ is still too low to allow an appreciation of the
convergence of the derivative expansion. Instead of presently producing one
\textquotedblleft best\textquotedblright\ estimate for each critical
exponent at a given value of $N$, it is preferable to maintain the freedom of $b$ in
order to better emphasize the early beginnings of some criteria of
convergence if any (see section \ref{Comp}).

Tables \ref{Tab1} and \ref{Tab2} display the estimates of $\eta $%
, $\nu $ and $\omega $ as obtained for four values of $b$ and $N$
varying from 1 to 20 (while $d=3$).
 
For $N=1$, the results of reference  \cite{5744} (preferred value of
$\eta^{\mathrm{opt}}$ for $b=0.11$) is obtained again. In this latter work
 I had proposed a criterium of choice of the value of $%
b$ which gave a preferred value of $\eta $. It was based on the idea of
Golner \cite{212} of a global minimization of the magnitude of the function $%
Z\left( \phi \right) $ (i.e., $\mathrm{v}_{2}\left( \phi \right) $). However
the extension of this criterion to $N>1$ is not easy to implement because
the function $Z\left( \phi \right) $ has been split into two parts ($Z(\rho)$ and $%
Y(\rho)$).

The further discussion of these results is left to section \ref{Comp}.

Table \ref{Tab3} displays a comparison between the LPA and $O(\partial^2)$ results for $d=3$, $N=1$ and $b=0.11$ of the series of subcritical exponents $\omega_n$
(with $\omega_1=\omega$). 
Compared to LPA, the order $O\left( \partial ^{2}\right) $ has increased
the potential number of subcritical exponents (due to the supplementary terms in the truncated action
considered). For instance, if one adopts the dimensional analysis of perturbation theory, then
 at order LPA a term like $\phi^6$ contributing to $S[\phi]$ 
 is associated with the subcritical $\omega_2$ while a $\phi^8$-term generates $\omega_3$. But at order
$O\left( \partial ^{2}\right) $, new terms involving two derivatives contribute to  $S[\phi]$, and 
a term proportionnal to $\phi^2(\partial \phi)^2$ induces a priori an 
in-between correction exponent. Despite important differences between the classical dimensions of the couples
  $\phi^6$ and $\phi^2(\partial \phi)^2$, table \ref{Tab3} clearly shows
 that, for $N=1$, the order $\partial^2$ induces a splitting of the LPA values of the subcritical exponents $\omega_n$ for $n>1$ ($\omega$ is,
in the perturbative approach, 
associated to the unique $\phi^4$ coupling and is not, fortunately for the perturbative framework, subject to this splitting).
This simple splitting is presumably not preserved for other values of $N$ due to the supplementary contribution of $Y(\phi)$.

\subsection{Comparison with other studies, discussions and conclusion}
\label{Comp}

\begin{figure}
\includegraphics[height=150mm]{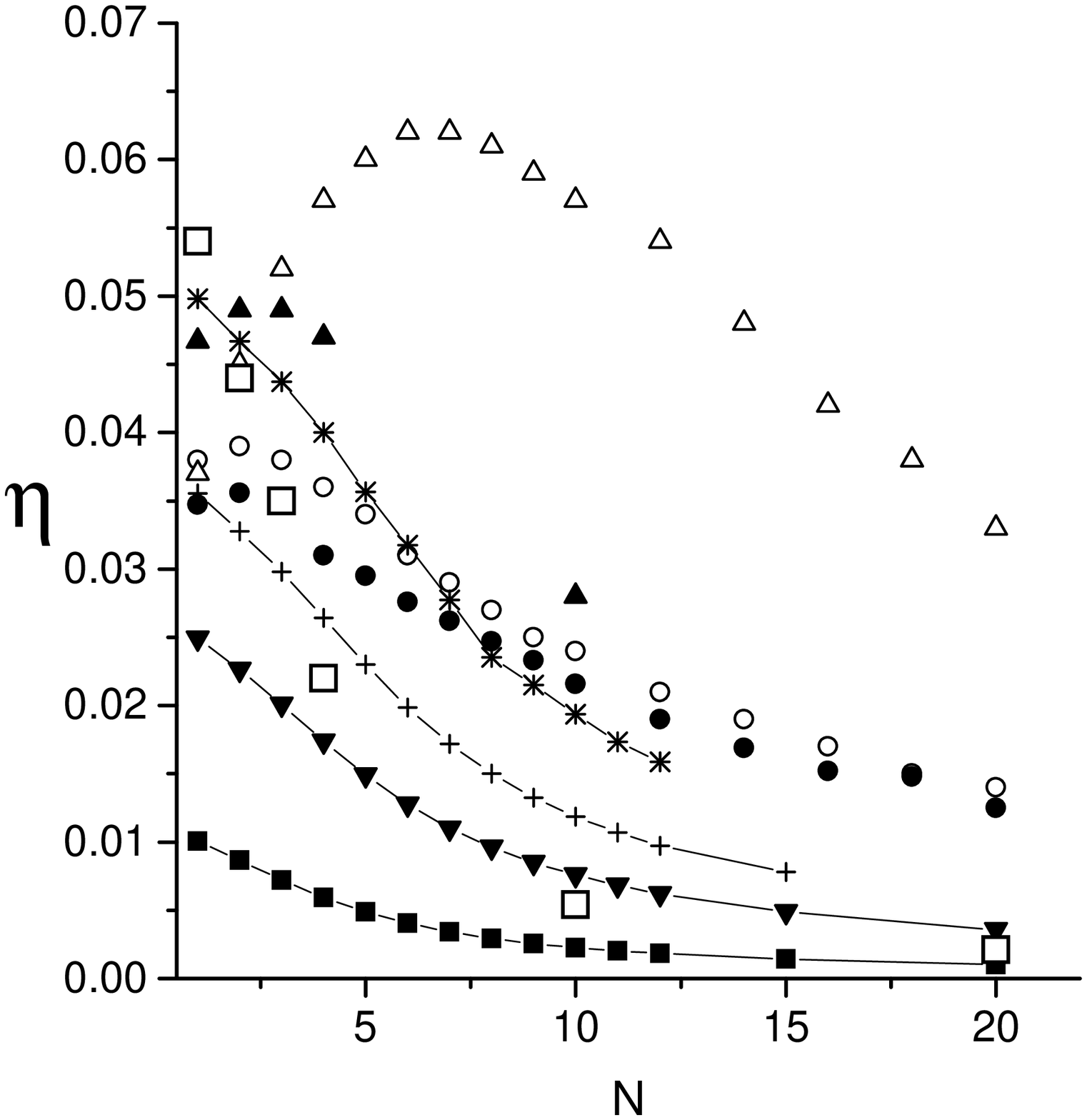}
\caption{\label{Fig1} Critical exponent $\eta$ as function of $N$  ($d=3$).  Open and full circles 
represent the standards, they were obtained from the resummation of six (\opencircle from \cite{4006})
or seven  (\fullcircle from \cite{4228})
orders of the perturbation field
theory. Recent estimates of \cite{5702}
from the same perturbative series at order three only are also represented (\opentriangle). The other
estimates are from the exact RG equation expanded up to $O\left( \partial ^{2}\right) $ in the derivative expansion: 
full triangle up from \cite{4754}, \opensquare\   
from \cite{3828}. The points linked by straight lines are from the present work: \fullsquare\   for $b=0.03$,
full triangle down for $b=0.11$, + for $b=0.17011$, * for b=0.25611. See text for discussion.}
\end{figure}

Figure \ref{Fig1} shows the evolution of $\eta$ with $N$ from different works. The results obtained from the resummation
of six \cite{4006} and seven \cite{4228} orders of the perturbation field theory serves the purpose of standards 
(other accurate estimates of the critical exponents, especially for $N\leq4$, exist in the litterature, 
for a review see \cite{4948}). One sees that
the present study
yields generally small values of $\eta$ compared to the standards except for small $N$ and for the highest values of $b$. 
The evolution of $\eta$ with $N$ is smoother than in the work of Morris and Turner \cite{3828} but, as in this latter work,
 the non-monotonic behavior
of the standards (responsible for the maximum of $\eta$ about $N=2$ or $3$) is not reproduced. Instead, the results of Gersdorff 
and Wetterich \cite{4754} are better. The present results are however not so bad if one keeps in mind that the first estimate
of $\eta$ in the derivative expansion is given at order $\partial^2$. In particular, figure \ref{Fig1} shows also recent estimates from the resummation
of three orders of the 
perturbative series using an efficient  method \cite{5702}. One sees that the present estimates withstand the comparison (except
the monotonic evolution with $N$). Notice also, for $N$ fixed, the monotonic evolution of $\eta$ with $b$ already mentioned.

\begin{figure}
\includegraphics[height=150mm]{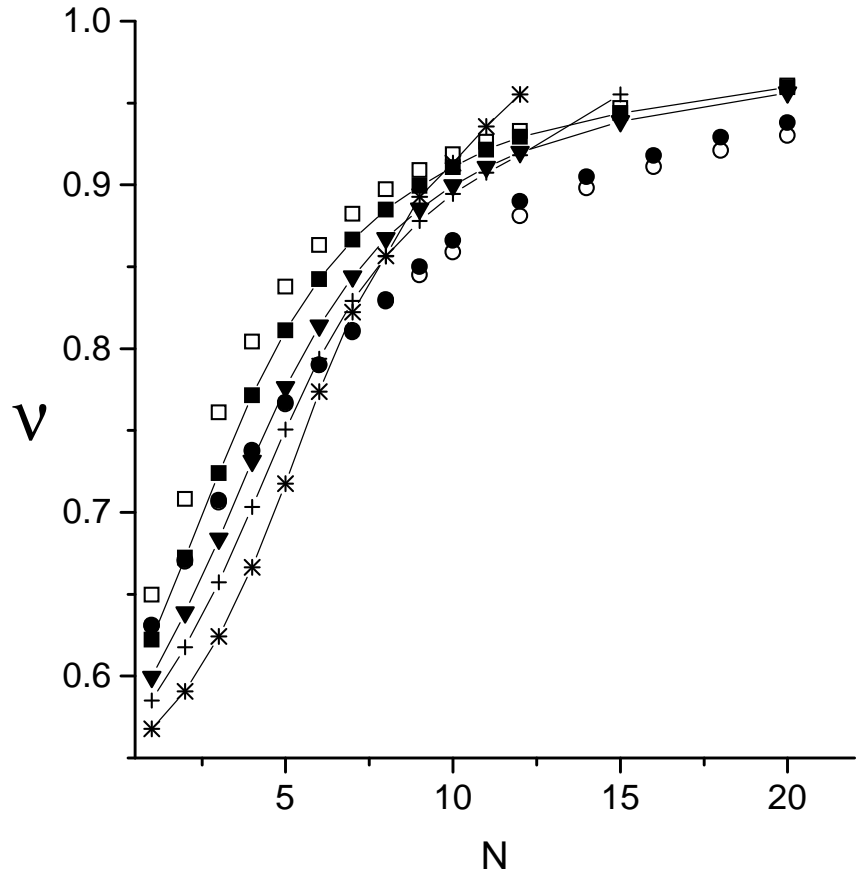}
\caption{\label{Fig2} Critical exponent $\nu$ as function of $N$ ($d=3$).  Open and full circles 
represent the standards as in figure \ref{Fig1}.  The other
estimates are from the present work:  \opensquare\  from order LPA, the points linked by straight lines 
are from  $O\left( \partial ^{2}\right) $ in the derivative expansion:  \fullsquare\   for $b=0.03$,
full triangle down for $b=0.11$, + for $b=0.17011$, * for b=0.25611. See text for discussion.}
\end{figure}

The results for the critical exponent $\nu$ are more interesting to discuss because the order $\partial^2$ provides
its second estimate. Figure \ref{Fig2} shows the evolution with $N$ of $\nu$ at order $\partial^2$ compared to the results 
at order LPA (obtained in the present work), the standards \cite{4006,4228} are reproduced also. 
Again, one observes that the results for large $N$ are not as good as 
for small values. However, for these latter values, one clearly sees that there is a range of values of $b$ where the two estimates
at orders
LPA and $\partial^2$ flank the standards and another range where the two present estimates are on the same side (with respect to the 
standards). This is a phenomenon often observed in convergent series the elements of which depend on a free parameter (like $b$) but the
resumed series does not: on varying the free parameter, one may observe monotonic or alternate approaches to the limit. 
These features may be 
used to determine error bars. Presently one additionnal order would be necessary to propose such error bars. Figure \ref{Fig2} shows 
also that, when $N$ increases the dependence of $\nu$ on $b$ becomes non-monotonic. This is interesting since such extrema may indicate
a vestige of the primarily independence on $b$ of the exact RG equation. Why this effect does not occur at small values of
$N$ is not explained. Once more, at least one supplementary order of the derivative expansion would be necessary to understand this point.

\begin{figure}
\includegraphics[height=150mm]{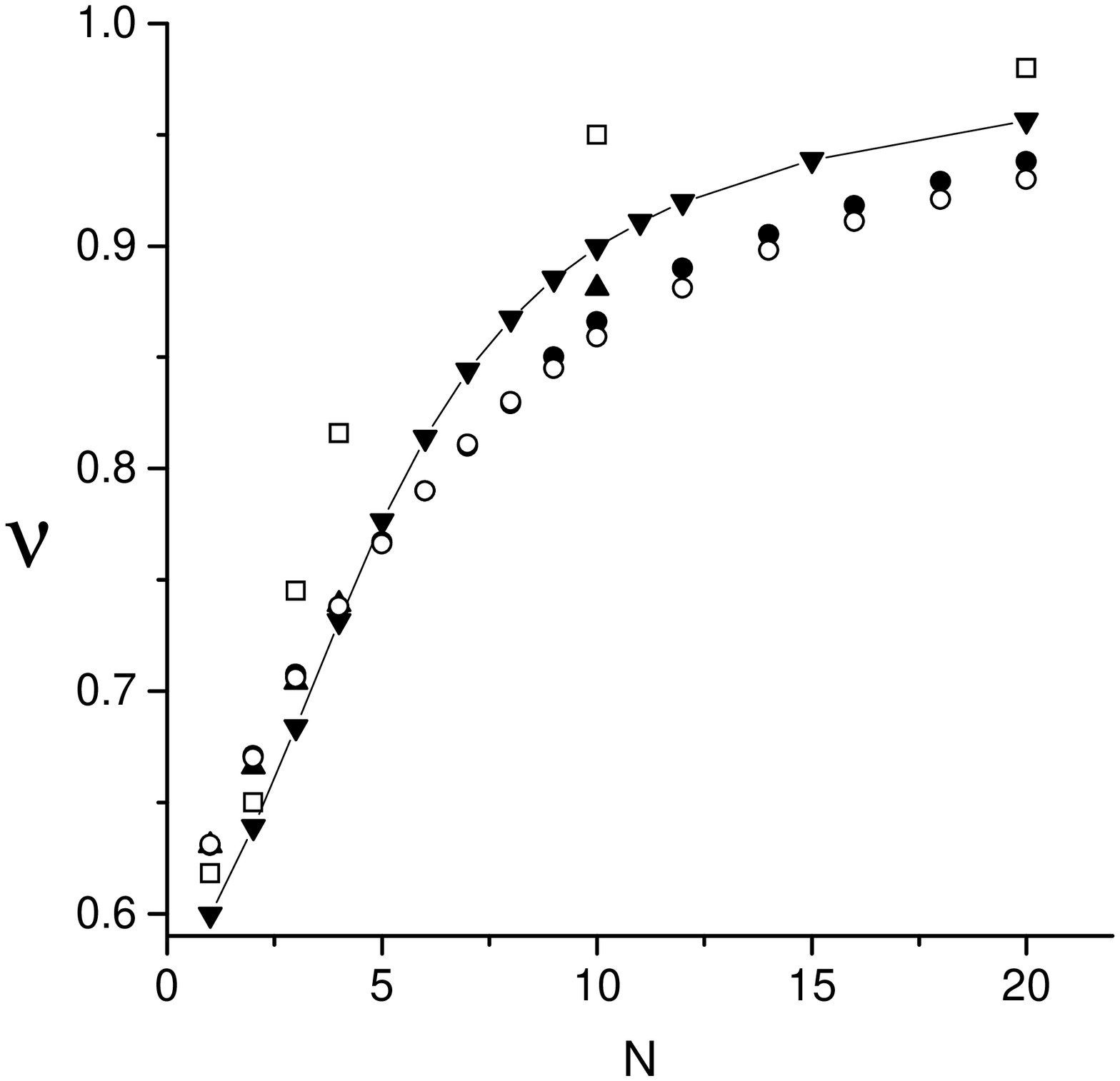}
\caption{\label{Fig3} Critical exponent $\nu$ as function of $N$ ($d=3$).  Open and full circles 
represent the standards as in figure \ref{Fig1}.  The other
estimates are from the exact RG equation expanded up to $O\left( \partial ^{2}\right) $ in the derivative expansion: 
full triangle up from \cite{4754}, \opensquare\   
from \cite{3828}. The estimates of the present study for $b=0.11$ is reported (full triangles down linked by straight lines,
 see figure \ref{Fig2} for the other values of $b$). See text for discussion.}
\end{figure}

Figure \ref{Fig3} shows the results for $\nu$ coming from \cite{3828} and \cite{4754} compared to the present results for $b=0.11$ and the 
standards  \cite{4006,4228}. One observes that the present results are globally better than in \cite{3828} and that again the estimates
of Gersdorff 
and Wetterich \cite{4754} are excellent (for small $N$ the points almost coincide with the standards).

\begin{figure}
\includegraphics[height=150mm]{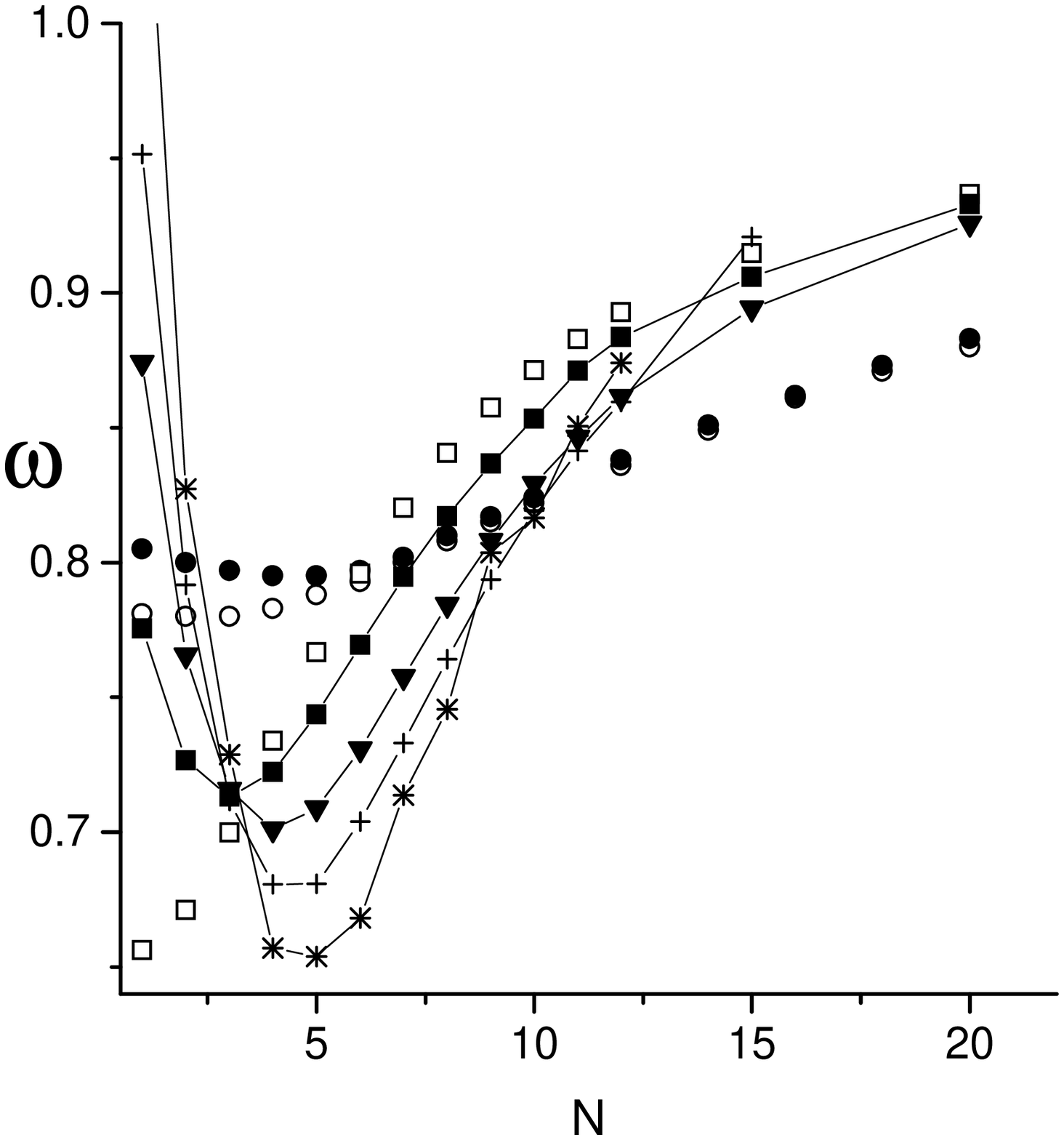}
\caption{\label{Fig4} Subcritical exponent $\omega$ as function of $N$ ($d=3$).  Open and full circles 
represent the standards as in figure \ref{Fig1} but the opencircles (\opencircle) are from \cite{3923}.  The other
estimates are from the present work:  \opensquare\  from order LPA, the points linked by straight lines 
are from  $O\left( \partial ^{2}\right) $ in the derivative expansion:  \fullsquare\   for $b=0.03$,
full triangle down for $b=0.11$, + for $b=0.17011$, * for b=0.25611. See text for discussion.}
\end{figure}

Figure \ref{Fig4} shows the present results for $\omega$ at order LPA and $\partial^2$ compared to the standards \cite{3923,4228}. 
This figure is the matching piece to figure \ref{Fig2} and the same kinds of remarks stand: monotonic and alternate approaches to the 
standards at fixed $N$ exist as well as non-monotonic dependences on $b$. The magnitude of these effects are larger than for $\nu$ 
and the accuracy is worse, but this is expected for a subleading eigenvalue: the accuracy decreases as the order of the 
eigenvalue increases.

\begin{figure}
\includegraphics[height=150mm]{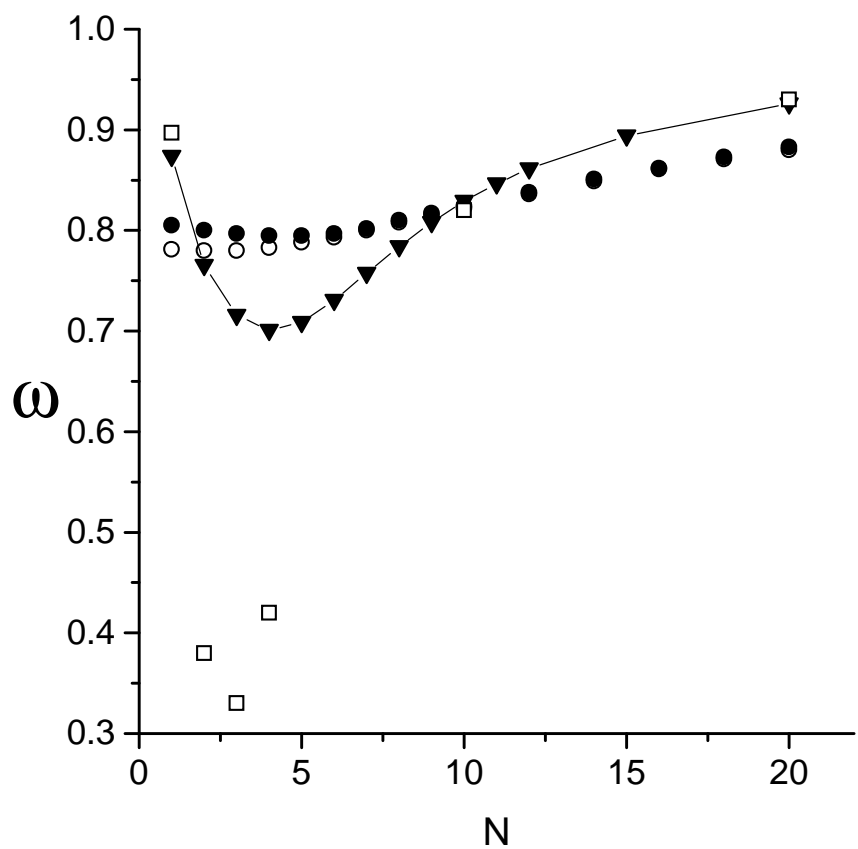}
\caption{\label{Fig5} Subcritical exponent $\omega$ as function of $N$ ($d=3$).  Open and full circles 
represent the standards as in figure \ref{Fig4}.  The other
estimates are from the exact RG equation expanded up to $O\left( \partial ^{2}\right) $ in the derivative expansion: 
 \opensquare\   
from \cite{3828}. The estimates of the present study for $b=0.11$ is reported (full triangles down linked by straight lines,
 see figure \ref{Fig4} for the other values of $b$). See text for discussion.}
\end{figure}

Figure \ref{Fig5} shows the results for $\omega$ coming from \cite{3828} compared to the present results for $b=0.11$ and the 
standards  \cite{3923,4228}. One observes that the present results are much better than in \cite{3828}. 

One may regret that
Gersdorff 
and Wetterich \cite{4754}, who obtained excellent values for $\nu$ and $\eta$, had not estimated $\omega$.  As already said,
this study \cite{4754} does not follow the conventional approach defined in the introduction. In particular nothing is said on
the way the reparametrization invariance is accounted for. In fact, instead of
leaving free the value of $Z(0)=Z_0$ to get a function $\eta(Z_0)$, the procedure followed in \cite{4754} was to attach the 
determination
of  $\eta$ to the minimum of the potential. This condition fixes $Z_0$ and the arbitrariness carried by the reparametrization
invariance is implicitly removed this way. Because
Morris and Turner \cite{3828} have considered an equation of the same family  and have obtained disappointing results, 
I think that this particular way of choosing $\eta^{\mathrm{opt}}$ could be the main
reason for the excellent estimates of the critical exponents obtained in \cite{4754}. It will be interesting to adapt it to the study
of the 
Wilson-Polchinski RG equation.

To conclude, the derivative expansion at order $O(\partial^2)$ already displays a tendency to converge. This must be confirmed by 
considering the next order which is in progress \cite{progress}. The study of reference \cite{5469} which for $N=1$ follows the procedure of reference 
\cite{4754} and the optimization process of reference \cite{4714} is very
encouraging. I think that the Wilson-Polchinski RG equation, which is the simplest exact RG equation, is better adapted to the 
estimation of the
critical exponents. Further studies should be undertaken with a view to better determine the status of the derivative expansion.

\ack I am indebted to C. Bagnuls for discussions and
encouragements all along this work.

\end{document}